\documentclass[aps,prb,twocolumn,superscriptaddress,amsmath,amssymb,showpacs]{revtex4-1}
\usepackage{amssymb}
\usepackage{amsmath}
\usepackage{amsfonts}
\usepackage{ dsfont }
\usepackage{graphicx}
\usepackage{bm}
\usepackage{subfigure}
\usepackage{epsfig}
\usepackage{color}
\usepackage{bbding}
\usepackage{array,multirow}
\usepackage{soul}
\usepackage[ansinew]{inputenc}

\usepackage{youngtab}
\usepackage{young}

\def\bk{{\bold{k}}}
\def\br{{\bold{r}}}

\begin{document}

\title{The Symplectic Fermi Liquid and its realization in cold atomic systems}

\author{Aline Ramires}
\affiliation{Institute for Theoretical Studies, ETH Zurich, CH-8092, Zurich, Switzerland.}

\begin{abstract}
In this work we study a system of interacting fermions with large spin and SP(N) symmetry. We contrast their behaviour with the case of SU(N) symmetry by analysing the conserved quantities and the dynamics in each case. We also develop the Fermi liquid theory for fermions with SP(N) symmetry. We find that the effective mass and inverse compressibility are always enhanced in the presence of interactions, and that the N-dependence of the enhancement is qualitatively different in distinct parameter regimes. The Wilson ratio can  be enhanced, indicating that the system can be made closer to a magnetic instability, in contrast to the SU(N) scenario. We conclude discussing what are the experimental routes to SP(N) symmetry within cold atoms and the exciting possibility to realize physics in higher dimensions in these systems.
\end{abstract}

\date{\today}

\maketitle

\section{Introduction}


Symmetries have always played an important role in physics. Some can occur naturally, being intrinsic to the system of interest, others require fine-tuning of parameters in order to be present. The experimental control we have over cold atoms allow us to construct systems with symmetries which are larger than the ones naturally present in matter. The study of systems with enlarged symmetries is in principle very attractive: these have larger degeneracies, which can be used to the experimentalists' advantage in adiabatic cooling in order to achieve low temperatures more effectively \cite{Geo, Bla, Hof}; also, the enhancement of quantum fluctuations and the presence of more degrees of freedom provide theorists with the possibility to study new phases of matter \cite{Wu10}. 

The realization of SU(N) symmetry has already been explored both theoretically and experimentally. This symmetry has been identified in alkaline-earth and Yb cold atomic systems \cite{Caz,Gor}. These atoms have a completely full outer electronic shell of s-character, so total electronic spin and angular momentum equal to zero. As a consequence, the nuclear spin is effectively decoupled from the electronic degrees of freedom and the s-wave scattering lengths for all nuclear spin configurations are equal. The effective Hamiltonian describing the interacting alkaline-earth atoms is SU(N) symmetric, where $N=2f+1$ and $f$ is the hyperfine spin (which for these atoms is equal to the nuclear spin). Experimental realizations of systems with SU(N) symmetry were already reported, with ultracold Yb isotopes trapped in one dimension \cite{Pag} or loaded in a 3D lattice \cite{Tai}. Theoretically several aspects of the SU(N) symmetry were already explored, including the characterization of the Fermi Liquid behaviour \cite{Yip, Caz09}, magnetism \cite{Che,Kat,Her, Xu10,Gua}, superconductivity \cite{Ho,Cap}, multipolar orders \cite{Tu}, staggered flux order \cite{Hon} and topology \cite{Szi, Yan, Aro}.

The presence of SU(N) symmetry is restricted to alkaline-earth atoms with a non-zero nuclear spin, what in principle gives us a small number of options amongst all isotopes available in nature. The question that follows is: are there different enlarged symmetries which can be realized with other atomic isotopes? The answer is yes, and SP(N) symmetry is a good candidate since it is a subgroup of SU(N), therefore less restrictive. An early work by Wu et al. \cite{Wu03} on cold atoms with hyperfine spin $f=3/2$ explore a \emph{hidden SP(4) symmetry}. More recently\cite{Wu05} the presence of symplectic symmetry in higher-spin Hubbard models within cold fermions was discussed, and it is pointed out that the symplectic symmetry does not require any fine tuning for $f=3/2$, while higher hyperfine spin systems require certain tuning of the interactions for the symmetry to be present. Another subgroup of SU(N) is SO(N), which can also be realized in cold atomic systems with bosonic isotopes\cite{Jia}.

The main aspect that makes cold atom systems with SP(N) symmetry distinct from the ones with SU(N) symmetry is related to the dynamics of each spin component. In systems with  SU(N) symmetry one can understand each spin component as a different colour or flavour, and the interactions allow only for colour-preserving scattering, as depicted in  Fig.~\ref{SUNSPN} a) and c). On the other hand, in case of SP(N) symmetry, it is more intuitive to label the spin components with a colour and an arrow, which can be either up or down. The colour can be understood as the different magnitudes of the spin component, and the up and down arrows as their sign. The form of the interactions in this case allows for a very special kind of scattering, which takes a pair of atoms with the same colour (up and down) and transmute it to a pair of a different colour, as shown  in  Fig.~\ref{SUNSPN} b) and e). These points are reviewed and discussed in detail in Section \ref{SPN}. More generally, one would have spin-flip scattering processes which do not preserve each nuclear spin component but only the total angular momentum of the colliding atoms. This has been observed and controlled experimentally with the long-lived alkaline radioisotope $^{40} K$ in an optical lattice \cite{Kra}. 

SP(N) symmetry can only be realized with fermions ($N=2f+1$ can only acquire even values for the symplectic group, requiring half-integer hyperfine spins). At low enough temperatures, fermions can reach quantum degeneracy and behave as a Fermi liquid (FL). Fermi liquid behaviour is ubiquitous in condensed matter systems and a very robust state of matter. The original FL theory was developed for spin-1/2 fermions \cite{Lif} and recently there was a generalization for fermions with larger spin and SU(N) symmetry \cite{Yip}. In Section \ref{FL} we develop the Fermi liquid theory for SP(N) cold fermions, analyzing the effective mass, compressibility and susceptibility and contrast these results with the SU(N) Fermi liquid \cite{Yip}. The most interesting aspect of the analysis concerns magnetism. We distinguish between two kinds of susceptibilities: a generalized and a physical susceptibility. Both are renormalized in the same fashion in the presence of interactions and can be either enhanced or suppressed, depending on the parameter regime. In Section \ref{Con} we discuss the possible routes to realize SP(N) symmetry within cold atomic systems, and highlight exciting directions for future work which allows us to explore experimentally issues only thought to be in the theoretical realm, as physics in higher dimensions.


\section{SU(N) and SP(N) Symmetries in Cold Atoms}\label{SUSPN}

We start with a general model for cold atoms with hyperfine-spin $f$. We assume a dilute gas with contact interactions so at low energies only the s-wave scattering channel is relevant \cite{Lee, Caz}. We can write the effective Hamiltonian as:
\begin{eqnarray}
H= H_0 + H_{I},
\end{eqnarray}
where $H_0$ is the kinetic part:
\begin{eqnarray}
H_0 = \int_\br \sum_{\alpha=-f}^f \Psi^\dagger_\alpha(\br) \left( -\frac{1}{2m}\nabla^2 + V (\br)  \right) \Psi_\alpha (\br),
\end{eqnarray}
which describes moving atoms under a trapping potential $V(\br)$. Here $\Psi_\alpha^\dagger (\br)$ and $\Psi_\alpha (\br)$ are creation and annihilation operators, respectively, for atoms with hyperfine spin component $\alpha$ located at $\br$ which at this stage can be either bosons or fermions. The interacting part reads:
\begin{eqnarray}
H_{I} = \frac{1}{2} \int_\br \sum_{\alpha,\beta,\mu,\nu=-f}^f  \!\!\! \Psi_\beta^\dagger(\br) \Psi_\alpha^\dagger (\br) \Gamma_{\alpha\beta;\mu\nu} \Psi_\mu(\br) \Psi_\nu (\br),
\end{eqnarray}
where the interaction vertex can be decomposed in different total angular momentum channels as\cite{Ho98}:
\begin{eqnarray}\label{IntVer}
\Gamma_{\alpha\beta;\mu\nu} = \sum_{F=0}^{2f } g_F \sum_{M=-F}^F \langle f \alpha, f \beta | F M\rangle \langle F M | f \mu, f \nu\rangle.
\end{eqnarray}
Here $F$ is the total angular momentum of the two interacting atoms, $M$ its component and $ \langle f \alpha, f \beta | F M\rangle$ are Clebsch-Gordan coefficients (CGC). $g_F$ is the strength of the interaction in the channel with total angular momentum $F$. The model could similarly be written for atoms in an optical lattice, and the discussion below, concerning symmetries, should follow in an analogous fashion.


It can be shown by analyzing $H_I$, that only even-$F$ channels contribute to scattering. One can take $\alpha \leftrightarrow \beta$, use the properties of the CGC shown in Appendix \ref{AppCGC} and the fact that the (fermionic) bosonic operators (anti-) commute to rewrite the interaction term explicitly as:
\begin{eqnarray}
H_{I} &=& \frac{1}{4} \int_\br \sum_{\alpha,\beta,\mu,\nu=-f}^f \sum_F g_F (1+\eta  (-1)^{2f-F}) \\ \nonumber &\times& \sum_M  \Psi_\beta^\dagger  \Psi_\alpha^\dagger  \langle f \alpha, f \beta | F M\rangle \langle F M | f \mu, f \nu\rangle \Psi_\mu  \Psi_\nu  
\end{eqnarray}
where $\eta=+1$ for bosons and $\eta=-1$ for fermions. Note that for either bosons or fermions the factor $(1+\eta  (-1)^{2f-F})$ simplifies to $(1+ (-1)^{F})$, which is zero for odd-F and equal to 2 for even-F. This is a consequence of the compensation of the factors $\eta$ and $(-1)^{2f}$, which product is always equal to one since $f$ is an integer for bosons and a half-integer for fermions. Out of the total $2f + 1$ scattering channels, only $f+1$ for bosons or $f+1/2$ for fermions actually contribute to scattering.

\subsection{SU(N) symmetry}\label{SUN}

We start the discussion towards SP(N) symmetry showing  first that SU(N) can be realized in this system under the special condition of $g_F=g$, meaning that the interactions in all scattering channels are the same. In order to prove the presence of the symmetry, we can evaluate the commutator of the generators of SU(N) group with the Hamiltonian. SU(N) has $N^2-1$ generators, which can be written as:
\begin{eqnarray}\label{SUGen}
O_{\alpha\beta} = \int_\br \Psi_\alpha^\dagger (\br) \Psi_\beta (\br),
\end{eqnarray}
where each index $\alpha$ and $\beta$ can run over $N=2f+1$ values. Note that not all the generators are linearly independent since the Casimir operator $C=\sum_\alpha O_{\alpha\alpha}$ is a constant. These generators follow the SU(N) commutation relation:
\begin{eqnarray}
[O_{\alpha\beta}, O_{\mu\nu}] = O_{\alpha\nu} \delta_{\mu\beta} - O_{\mu\beta} \delta_{\alpha\nu}.
\end{eqnarray}

The commutator with the non-interacting part of the Hamiltonian is rather trivial and equal to zero. Evaluating now the commutator with the interacting part, we find after some manipulation:
\begin{eqnarray}\label{HIO}
\left[H_I,  O_{\alpha'\beta' } \right]  &=&\int_\br \Big( \Gamma_{\alpha\beta;\mu \alpha' } \Psi_\beta^\dagger (\br) \Psi_\alpha^\dagger (\br) \Psi_\mu (\br)  \Psi_{\beta ' }(\br)  \\ \nonumber  && -  \Gamma_{\alpha \beta '  ;\beta \mu} \Psi_\beta^\dagger (\br) \Psi_{\alpha' }^\dagger (\br) \Psi_\alpha (\br) \Psi_\mu (\br) \Big),
\end{eqnarray}
which is generally not equal to zero. In the equation above the sum over repeated indexes is implied. Under the consideration that the interactions in all scattering channels are equal, $g_F=g$, we can use the orthogonality condition of the CGC (Eq.~\ref{CGCOrt1}) to simplify the interaction vertex to:
\begin{eqnarray}
\Gamma_{\alpha\beta;\mu\nu}^{SU(N)} = g  \delta_{\alpha\mu}\delta_{\beta\nu},
\end{eqnarray}
so we can write the explicit form of the interaction part of the Hamiltonian in case of SU(N) symmetry:
\begin{eqnarray}\label{SUInt}
H_{I}^{SU(N)} = \frac{g}{2} \int_\br \sum_{\alpha,\beta=-f}^f  \!\!\! \Psi_\alpha^\dagger(\br) \Psi_\beta^\dagger (\br) \Psi_\beta(\br) \Psi_\alpha (\br).
\end{eqnarray}

It is a simple task to show that
\begin{eqnarray}\label{HIO}
\left[H_{I}^{SU(N)},  O_{\alpha'\beta' } \right]  &=&0,
\end{eqnarray}
what proves the SU(N) symmetry. Note that this result is independent of the bosonic or fermionic character of the atoms.

The diagonal generators:
\begin{eqnarray}
O_{\alpha\alpha} = \int_\br \Psi_\alpha^\dagger (\br) \Psi_\alpha (\br) =n_\alpha
\end{eqnarray}
commute with the Hamiltonian, so these are conserved quantities. The total number of particles with a given spin-component, or flavour, is preserved if the interaction, is the same for all channels and SU(N) symmetry is realised.

\subsection{SP(N) symmetry}\label{SPN}


Given the discussion above, now we move to the study of the SP(N) case. One subtlety about the SP(N) generalization is that the group is only defined for even-N and its realization is possible only with fermionic atoms. We can check the presence of SP(N) symmetry by evaluating the commutator of the generators of SP(N) with the Hamiltonian. SP(N) is a subgroup of SU(N) and has $N(N+1)/2$ generators, which can be written as specific linear combinations of the SU(N) generators defined above in Eq.~\ref{SUGen}:
\begin{eqnarray}\label{SPGen}
S_{\alpha\beta} &=& O_{\alpha\beta} +(-1)^{\alpha+\beta} O_{-\beta-\alpha},
\end{eqnarray}
where again $\alpha$ and $\beta$ can run over $N$ values. Note that these generators are  not all linearly independent given the relation $S_{\alpha\beta} = (-1)^{\alpha+\beta}S_{-\beta-\alpha}$.

The commutator with the non-interacting part of the Hamiltonian is again trivial. Concerning the interacting part, given that SP(N) is a subgroup of SU(N), if a Hamiltonian has SU(N) symmetry (if $g_F=g$), it will also commute with the generators of SP(N). Note, though, that this is not what we are looking for since the actual symmetry of the system is still SU(N) in this case. We need to look for a way to break the full SU(N) symmetry down to SP(N). From the strongly correlated systems perspective, it is known that SP(N) was introduced in order to deal with valence bonds in frustrated magnetism \cite{Rea3,Fli09} and singlet pairing \cite{Fli08,Fli12}. It is suggestive then, that the zero total angular momentum channel $g_{F=0}$ is the important one to distinguish SU(N) from SP(N) symmetry.

We can use the results obtained for SU(N), before the assumption that all channels have the same interaction strength, given by Eq.~\ref{HIO}, and look at the less restrictive condition of having $g_0\neq g_{F>0}=g$. In this case we can combine all terms with the same magnitude of the interaction and part of $g_0$ to use the orthogonality condition of the CGC, leading to zero contribution to the commutator, as found in the SU(N) discussion above. We are left with a term proportional to $\Delta g=g_0-g$ in the $F=0$ channel to be evaluated. Now the interaction vertex simplifies to
\begin{eqnarray}
\Gamma_{\alpha\beta;\mu\nu}^{SP(N)}  = \Gamma_{\alpha\beta;\mu\nu}^{SU(N)} - \frac{\Delta g}{N} (-1)^{\alpha+\mu} \delta_{\alpha,-\beta} \delta_{\mu,-\nu}, \end{eqnarray}
after using Eq.~\ref{CGC0}, identifying $2f+1=N$, and remembering that we are dealing only with fermions, so $2f$ is always an odd number. Under these considerations the interacting part of the Hamiltonian for the case of SP(N) symmetry can be written explicitly as:
\begin{eqnarray}\label{SPInt}
&&H_{I}^{SP(N)} = \frac{g}{2} \int_\br \sum_{\alpha,\beta=-f}^f  \!\!\! \Psi_\alpha^\dagger(\br) \Psi_\beta^\dagger (\br) \Psi_\beta(\br) \Psi_\alpha (\br)\\ \nonumber
&+&\frac{\Delta g}{2N} \int_\br \sum_{\alpha,\beta=-f}^f  \!\!\! (-1)^{\alpha+\beta}\Psi_\alpha^\dagger(\br) \Psi_{-\alpha}^\dagger (\br) \Psi_\beta(\br) \Psi_{-\beta} (\br).
\end{eqnarray}
Note that the first term is the same as the one present in the case of SU(N) symmetry. The second term, proportional to the detuning of the $F=0$ channel, is the part of the interaction which breaks SU(N) down to SP(N).

We can now evaluate the commutator of the interacting part of the Hamiltonian under the condition $g_0\neq g_{F>0}=g$ with the SU(N) generators to find:
\begin{eqnarray}
\left[H_{I}^{SP(N)},  O_{\alpha'\beta'} \right]&=&- \int_\br \sum_{\alpha=-f}^f \frac{\Delta g}{2N} (-1)^{\alpha}\\
\nonumber &&\times\left[ (-1)^{\alpha'} \Psi_\alpha^\dagger (\br) \Psi_{-\alpha}^\dagger (\br)  \Psi_{-\alpha'} (\br)   \Psi_{\beta' } (\br) \right.  \\ \nonumber &&- \left. (-1)^{\beta' }\Psi_{-\beta' }^\dagger (\br)  \Psi_{\alpha' }^\dagger (\br) \Psi_\alpha (\br) \Psi_{-\alpha} (\br) \right],
\end{eqnarray}
what is generally not equal to zero. Note though that:
\begin{eqnarray}
\left[H_{I}^{SP(N)},  (-1)^{\alpha'+\beta'} O_{-\beta'-\alpha'} \right]
&=& -  \left[H_{I}^{SP(N)},  O_{\alpha'\beta'} \right]\nonumber \\ 
\end{eqnarray}
so for the SP(N) generators:
\begin{eqnarray}
\left[H_{I}^{SP(N)},  S_{\alpha'\beta'}\right] = 0,
\end{eqnarray}
what indicates that the model with the interactions satisfying $g_0\neq g_{F>0}=g$ has SP(N) symmetry and not the larger SU(N) symmetry.

The diagonal generators are now:
\begin{eqnarray}
S_{\alpha\alpha} &=& \int_\br \left( \Psi_\alpha^\dagger(\br) \Psi_\alpha(\br) -\Psi_{-\alpha}^\dagger(\br) \Psi_{-\alpha}(\br) \right)\\ \nonumber&=& n_\alpha-n_{-\alpha}=m_\alpha
\end{eqnarray}
and commute with the Hamiltonian, so these are conserved quantities. The magnetization for a given magnitude of the spin-component, or the colour-magnetization, is preserved if the interaction, is the same for all but the $F=0$ channel and SP(N) symmetry is realised.

\subsection{Comparison of SU(N) and SP(N) symmetric systems}

\begin{figure}[t]
\begin{center}
\includegraphics[width=\linewidth, keepaspectratio]{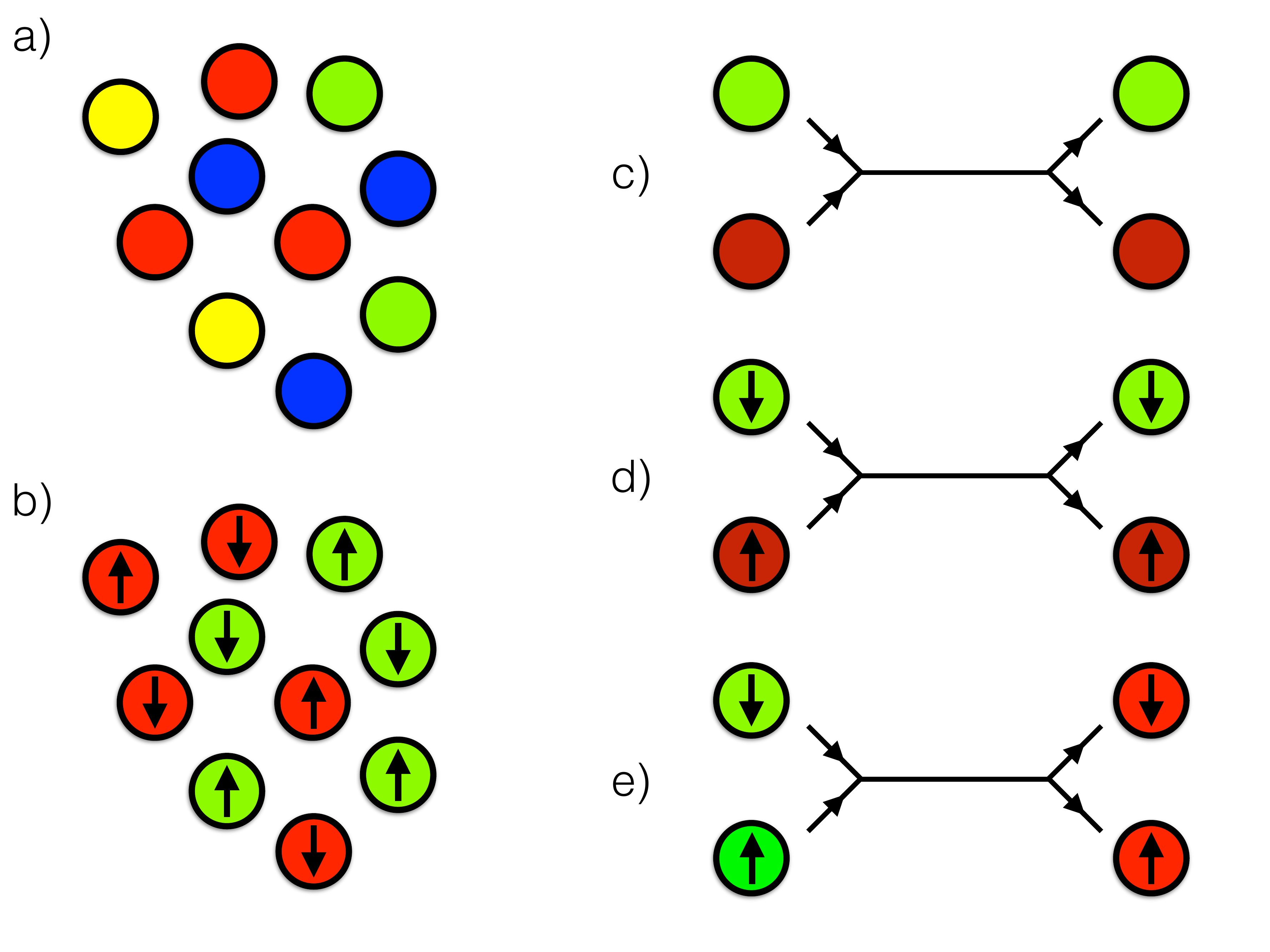}
\caption{Schematic comparison of SU(4) and SP(4) systems. An SU(4) system has 4 different flavours, represented by the 4 different colours in (a). In contrast, the SP(4) system represented in (b) has an extra label as ``up" or ``down" for two colours, in a total of four different flavours. In (c) we depict a SU(4) representative scattering process, which is colour-preserving. In (d) we represent the analogous scattering for SP(4). In the symplectic case there is also the possibility for processes as depicted in (e), where ``up" and ``down" pairs of fermions of a given colour can transmute into a pair of a different colour.} 
\label{SUNSPN}
\end{center}
\end{figure}

At this point a discussion on the physical implications of SU(N) and SP(N) symmetries is interesting. From the explicit form of the Hamiltonians in Eqs.~\ref{SUInt} and \ref{SPInt}, it becomes clear that in the first case two particles with components $\alpha$ and $\beta$ can only scatter into states with the same spin components, so the number of particles with each spin component is a constant. This is illustrated pictorially in Fig.~\ref{SUNSPN} c). On the other hand, the SP(N) interaction has a second contribution which allows the spin components, or colours, to change. Now two particles with the same colour and up and down arrows ($\beta$ and $-\beta$, for example) can scatter into a pair of states with opposite arrows and a different colour ($\alpha$ and $-\alpha$, for example). Fig.~\ref{SUNSPN} e) illustrates this point.

Another difference between SU(N) and SP(N) symmetries concerns the dynamics of the system. Given an initial state with a specific occupancy of the different spin components, the SU(N) and SP(N) cases will evolve differently. In particular, if we have alkaline-earth atoms with hyperfine spin $f$ and all the levels are occupied, the system has SU($N=2f+1$) symmetry. On the other hand, if only a certain subset $n<2f+1$ of the flavours is occupied, the symmetry which is realized is SU(n$<$N). This is a consequence of the fact that the number of particles in each flavour is conserved.  In contrast, in the symplectic scenario, the interaction allows for spin component transmutation. Even if we load the system with a single colour (with up and down arrows), the system will equilibrate to the lowest energy state with same occupation number to each flavour, and the symmetry is actually the maximal SP($N=2f+1$). Interestingly enough, if one initially traps only positive components, so they cannot pair with their complements in order to transmute into other flavours, then the symmetry is  lowered to SU($n\leq N/2$), where $n$ is the number of unpaired components trapped.

In the introduction it was already mentioned that SP(4) symmetry has been pointed out for systems with $f=3/2$ \cite{Wu03,Wu05}. What is special about $f=3/2$ is the fact that it naturally satisfies the condition for SP(4), without any fine-tuning.  We know that, by symmetry, only even channels contribute to scattering, so only $F=0,2$ are the allowed scattering channels for $f=3/2$. Even if the interactions in all channels are different, we fall under the condition with the $F=0$ channel different from all other channels (here only one, $F=2$). For larger spins we would have extra channels present, with different interactions. For instance, for $f=5/2$ there are three channels $F=0,2,4$. To realise SP(N) symmetry in this case one needs to tune the interactions for the $F=2$ and $F=4$ channels to be the same. A compilation of results for the particular case of SP(4) can be found in Wu \cite{Wu}, with more recent results on magnetism\cite{Kole, Xu09} and superconductivity \cite{Cap}. As we will discuss in more detail in Sec.~\ref{ExpRea}, unfortunately nature does not provide us with atoms with hyperfine spin-3/2 which have no dipole-dipole interactions and are not alkaline-earths (these realize the larger SU(4) and not SP(4) symmetry), so a smart experimental setup is necessary in order to realize even the first non-trivial case of SP(N) with $N=4$.

From the discussion above it is suggestive that the presence of new scattering processes that allow for spin-flip scattering (or colour transmutation) brings new aspects that should be investigated and the possibility of more interesting physics, mainly concerning magnetism, to be found. As a first exploration of these consequences, in this work we focus on the effects on the Fermi liquid behaviour.

\section{The Symplectic Fermi Liquid}\label{FL}

Given the motivation above for the realization of SP(N) symmetry within cold fermions, now we analyse the Fermi liquid behaviour of a system of fermions with symplectic symmetry. This is the state that would be accessible in experiments if the temperature is below quantum degeneracy, but not low enough so that order is able to develop. From this section on we will focus on the following Hamiltonian, already Fourier transformed to momentum space:
\begin{widetext}
\begin{eqnarray}\label{SPFL}
H_{FL}^{SP(N)} &=& \sum_\bk \sum_{\alpha} \Psi_{\bk\alpha}^\dagger\left(\frac{\bk^2}{2m}-\mu\right)\Psi_{\bk\alpha}
+ \frac{g}{2}  \sideset{}{' }\sum_{\{\bk\}} \sum_{\substack{\alpha,\beta\\ \alpha \neq \beta} } \Psi_{\bk_1,\alpha}^\dagger \Psi_{\bk_2\beta}^\dagger  \Psi_{\bk_3\beta} \Psi_{\bk_4\alpha} \\ \nonumber
&+& \frac{\Delta g}{2N}\sideset{}{' } \sum_{\{\bk\}}\sum_{ \alpha, \beta }  (-1)^{\alpha+\beta}\Psi_{\bk_1\alpha}^\dagger \Psi_{\bk_2-\alpha}^\dagger \Psi_{\bk_3\beta} \Psi_{\bk_4-\beta},
\end{eqnarray}
\end{widetext}
where $\Psi_{\bk\alpha}^\dagger$ ($\Psi_{\bk\alpha}$) creates (annihilates) a fermion with momentum $\bk$ and spin component $\alpha$, which can assume half-integer values between $-f$ and $f$. The first term describes free fermions with mass $m$ and chemical potential $\mu$. Here we ignore the trapping potential, assuming the fermions explore a region in space with an almost constant potential. The second term introduces part of the interactions which is also present in the SU(N) case (here we make explicit that the sum does not allow $\alpha=\beta$ since we are dealing with fermions). The third term introduces a new interaction vertex, which is particular to the SP(N) case, where $\Delta g=g_0-g$. The primed sum over $\{\bk\}=\bk_1,\bk_2,\bk_3,\bk_4$ indicate the sum over all momenta, subject to momentum conservation $\bk_1+\bk_2=\bk_3+\bk_4$.

We can construct a FL theory, following the lines of Lifshitz and Pitaevskii\cite{Lif}, treating the quasiparticle distribution function and the quasiparticle energy as $N \times N$ matrixes, in which each index corresponds to a spin component running from $-f$ to $f$, where $f$ is a half-integer number. In Yip et al.\cite{Yip}, the authors generalize the FL theory to SU(N) symmetry and compute the effective mass, magnetic susceptibility and compressibility in terms of the new Landau parameters. Here we will comment on the generalization to the symplectic case, and how physical quantities depend on the parameter $N$ in this new scenario.

The change in the quasiparticle energy $\delta \epsilon_{\alpha\beta} (\bk)$ due to an infinitesimal change in the quasiparticle distribution function $\delta n_{\alpha\beta} (\bk)$ can be written as:
\begin{eqnarray}
\delta \epsilon_{\alpha\beta} (\bk)=\sum_{\bk'}\sum_{\mu,\nu} f_{\alpha\mu,\beta\nu}(\bk,\bk') \delta n_{\nu\mu} (\bk'),
\end{eqnarray}
where $f_{\alpha\mu,\beta\nu}(\bk,\bk')$ is the interaction function\cite{Lif}. The specific form of the interaction function depends on the actual interactions between the particles and it will be worked out in Sec.~\ref{IntFun} below. Given SP(N) symmetry, we can parametrize the interaction function as follows:
\begin{eqnarray}\label{FLPar}
f_{\alpha\mu,\beta\nu}(\bk,\bk') &=& f_s(\bk,\bk') \delta_{\alpha\beta}\delta_{\mu\nu} + f_\epsilon(\bk,\bk') \epsilon_{\alpha\mu}\epsilon_{\beta\nu}\\ \nonumber&+& f_a (\bk,\bk') \sum_A \Gamma^A_{\alpha\beta}\Gamma^A_{\mu\nu}.
\end{eqnarray}
Note that, differently from the SU(N) case now we have three different parameters: $f_s(\bk,\bk')$, $f_a(\bk,\bk')$ and $f_\epsilon(\bk,\bk')$. This reflects the fact that there are three independent 4-indexed invariants under SP(N) transformations. This point is discussed in more detail in Appendix \ref{AppSym}. Here $\epsilon$ is an antisymmetric matrix, $\Gamma^A$ are the generators of the specific symmetry group. The label $A$ runs from $1$ to the total number of generators, which is equal to $N(N+1)/2$ for SP(N). The generators are traceless and we choose them to be normalized as $Tr[\Gamma^A \Gamma^B] = \delta_{AB}$. Note that the generators introduced in Section \ref{SPN} do not satisfy this condition, but in Appendix \ref{AppGen} we show these can be redefined such that this normalization holds, making the following calculations more straightforward.


\subsection{Effective Mass and Compressibility}\label{Sus}

The effective mass and compressibility for the SP(N) FL can be computed in the same fashion as for the SU(N) or SU(2) FL, so we simply state and comment on the results in this section. The effective mass reads:
\begin{eqnarray}
\frac{m^*}{m} = 1+ N \overline{F_s(\theta) \cos\theta},
\end{eqnarray}
where
\begin{eqnarray}
F_s(\theta) = \rho^*(E_f) \left( f_s(\theta)+\frac{1}{N} f_\epsilon(\theta) \right),
\end{eqnarray}
with $\theta$ the angle between $\bk$ and $\bk'$, which are at the Fermi surface. The overline denotes average over the solid angle. Here we introduce the density of states per spin component at the Fermi energy $\rho^*(E_f) = m^* k_f/(2\pi^2)$, with $k_f$ the Fermi momentum, defined from the total particle density $\rho_T=N_T/V = N\frac{k_f^3}{6 \pi^2}$.

Analogously, the inverse compressibility $u^2 = \frac{N_T}{m}\frac{d\mu}{dN_T}$ is modified in the presence of interactions as:
\begin{eqnarray}
\frac{u^{*2}}{u^2} = \frac{1+ N \overline{F_s(\theta)}}{1+ N \overline{F_s(\theta) \cos\theta}}.
\end{eqnarray}
These results are similar in form to the SU(N) results, in which case $f_\epsilon(\theta)=0$.

\subsection{The Generalized and the Physical Magnetic Susceptibilities}\label{Sus}

Now we would like to focus the discussion on the magnetic susceptibility. Here we will make a distinction between two kinds of susceptibility: a generalized susceptibility $\chi_G$, and a physical susceptibility $\chi_P$. 

For the generalized susceptibility we define a generalized magnetization with components $m^A$, associated with a generalized magnetic field with components $h^A$ which couple to the respective generator as $- \mu_B h^A \Gamma^A$. This is the natural generalization of the SU(2) case, in which there are 3 generators (the three Pauli matrices), each one coupling to one component of the magnetic field in 3-dimensional space as $- \mu_B h^i \cdot \sigma^i$. This is a formal definition, and it is what was evaluated for the SU(N) FL as a generalized susceptibility\cite{Yip}. 

We can perform a similar calculation, following Lifshitz and Pitaevskii\cite{Lif}, and evaluate the change in energy due to the presence of an external generalized magnetic field as follows:
\begin{eqnarray}
\delta \epsilon_{\alpha\beta}(\bk) &=& -\mu_B \sum_A h^A \Gamma^A_{\alpha\beta} \\ \nonumber&+& \sum_{\mu,\nu}\int d\tau' f_{\alpha\mu,\beta\nu} (\bk,\bk') \delta n_{\nu\mu}(\bk'),
\end{eqnarray}
where the first term accounts for the change in energy due to the presence of the field, while the second takes into account feedback effects due to interactions. We transformed the sum over $\bk'$ into an integral introducing $d\tau'  = \frac{d\bk' }{(2\pi)^3}$. We use the ansatz
\begin{eqnarray}
\delta \epsilon_{\alpha\beta}(\bk) &=& -\mu_B \frac{\gamma}{2}\sum_A h^A \Gamma^A_{\alpha\beta},
\end{eqnarray}
where $\gamma$ is a parameter to be determined self-consistently. Using the fact that the generators are traceless, normalized and follow the symplectic condition $\epsilon \Gamma^A \epsilon = (\Gamma^A)^T$, we find:
\begin{eqnarray}
\gamma = \frac{2}{1+ \overline{F_a(\theta)}},
\end{eqnarray}
where
\begin{eqnarray}
F_a(\theta) = \rho^*(E_f) \left( f_a(\theta)-f_\epsilon(\theta) \right).
\end{eqnarray}
Finally, the generalized susceptibility is defined as:
\begin{eqnarray}
\chi_G h^A = m^A = \mu_B \sum_{\alpha\beta} \int d\tau \Gamma^A_{\alpha\beta} \delta n_{\beta\alpha} (\bk),
\end{eqnarray}
and has the form:
\begin{eqnarray}
\chi_G  =  \frac{2 \mu_B^2 \rho^*(E_f)}{1+  \overline{F_a(\theta)} }.
\end{eqnarray}
Note that the non-interacting susceptibility computed in this fashion is independent of N. This is in fact the result for what was defined as the generalized susceptibility $\chi_G$, but it goes against the physical intuition that if we have a Fermi gas with many spin components, all susceptible to a magnetic field, the susceptibility should depend on the number of components. Also, this computation assumes the existence of a generalized magnetic field with as many components as generators, so $N(N+1)/2$ components. This suggests that one would need to be in higher spacial dimensions in order to realize it. We are going to comment further on this aspect in Sec~\ref{highd}. 

Based on this discrepancy we evaluate now what we call the physical susceptibility $\chi_P$. The physical point of view asks the following question: what happens when we apply actual magnetic field (assuming a 3-dimensional space) to a system with an enlarged symmetry? The standard estimation for the magnetization of a pair of spin components $\alpha$ and $-\alpha$ is $m_\alpha=2\alpha \mu_B (n_\alpha-n_{-\alpha})$, and one can approximate $n_\alpha-n_{-\alpha}\approx \rho(E_f) 2 \alpha g \mu_B h^z$, where $h^z$ is now a physical magnetic field chosen to be in the z-direction. The total magnetization can then be written as:
\begin{eqnarray}
m^z= \sum_\alpha m_\alpha &=& 4 \rho(E_f) g \mu_B^2  \sum_\alpha \alpha^2 h^z
\end{eqnarray}
so from $m^z=\chi_P h^z$ we can identify the physical susceptibility as:
\begin{eqnarray}
\chi_P&=&\frac{2 \mu_B^2 \rho^*(E_f)}{1+  \overline{F_a(\theta)} } \frac{N (N^2-1)}{6},
\end{eqnarray}
which now depends on $N$. Note that in the case $N=2$ we recover the known SU(2) result, with the last fraction equal to one and $f_\epsilon(\theta)=0$.  This result should be valid for both SU(N) and SP(N) Fermi liquids, which will differ on the specific form of the renormalization factors, which we treat explicitly below.



Note that in both cases, for the generalized or physical susceptibilities, the effects of interactions lead to the same renormalization:
\begin{eqnarray}
\left(\frac{\chi^*_{G,P}}{\chi_{G,P}}\right)^{-1}&=& \frac{1+   \overline{F_a(\theta)}}{1+ N \overline{F_s(\theta) \cos\theta}}.
\end{eqnarray}

\subsection{Explicit form of the interaction function}\label{IntFun}

The interaction function is defined as the second variation of the total energy with respect to occupation numbers:
\begin{eqnarray}
f_{\alpha\mu,\beta\nu}(\bk,\bk') = \frac{\delta^2 E}{\delta n_{\beta\alpha}(\bk)\delta n_{\nu\mu}(\bk')}.
\end{eqnarray}

If we approach the interacting problem perturbatively, starting from the Fermi gas as the non-interacting problem, we have that the ground state over which averages are going to be taken has only diagonal non-zero occupation numbers $\delta n_{\alpha\beta}(\bk) = \delta n_{\alpha\alpha}(\bk) \delta_{\alpha,\beta}\equiv \delta n _\alpha (\bk) \delta_{\alpha\beta}$, so the only non-zero interaction functions have the form:
\begin{eqnarray}
f_{\alpha\beta}(\bk,\bk') = \frac{\delta^2 E}{\delta n_{\alpha}(\bk)\delta n_{\beta}(\bk')},
\end{eqnarray}
where we defined $f_{\alpha\beta}(\bk,\bk') \equiv f_{\alpha\beta,\alpha\beta}(\bk,\bk')$.

In first order, the contribution of the interactions to the total energy can be written as:
\begin{eqnarray}
E^{(1)} &=& \sum_{\bk_1,\bk_2} \sum_{\alpha,\beta} n_{\alpha}(\bk_1) n_{\beta}(\bk_2) \\ \nonumber &&\times \Bigg[  \frac{g}{2}(1-\delta_{\alpha\beta})  +\frac{\Delta g}{2N}\delta_{\alpha,-\beta}\Bigg].
\end{eqnarray}

Note that the last term in the first order correction is not present in the SU(N) scenario since $\Delta g=g_0-g$ is zero in that case. This term appears with a factor of $1/N$, but for small values of $N$ and significant $\Delta g$ it can play an important role. The interaction function in first order in the interactions is then:
\begin{eqnarray}
f_{\alpha\beta}^{(1)}(\bk,\bk') = \frac{g}{2}(1-\delta_{\alpha\beta})  +\frac{\Delta g}{2N}\delta_{\alpha,-\beta}.
\end{eqnarray}

\begin{figure}[h]
\begin{center}
\includegraphics[width=0.75\linewidth, keepaspectratio]{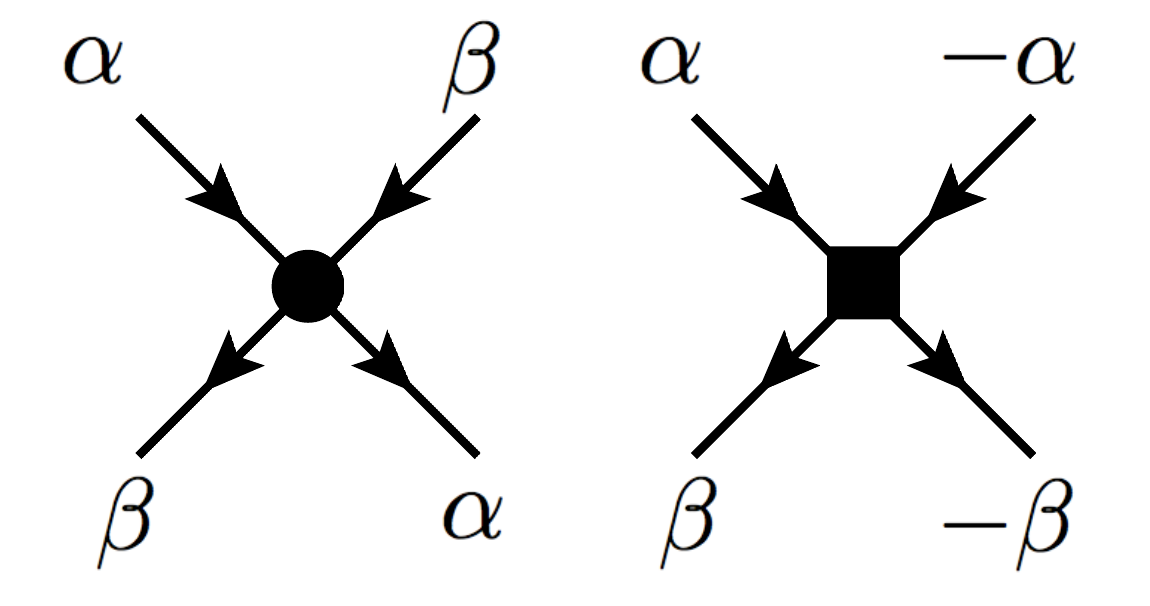}
\caption{Quasiparticle interaction vertices in first order in $g/2$ (circle) and in $\Delta g/2N$ (square), respectively, depicted as Feynman diagrams. Note that $\alpha\neq \beta$ in the first diagram.}
\label{FigVer}
\end{center}
\end{figure}

In second order, the contribution to the total energy is more involved. In favour of a concise notation we write:
\begin{eqnarray}
H_I^{SP(N)} =&&\sideset{}{' }\sum_{\{\bk\}}  \sum_{\alpha,\beta,\mu,\nu}  G_{\alpha\beta\mu\nu} \Psi^\dagger_{\bk_1\alpha} \Psi^\dagger_{\bk_2\beta} \Psi_{\bk_3\mu} \Psi_{\bk_4\nu},\nonumber\\
\end{eqnarray}
where
\begin{eqnarray}
G_{\alpha\beta\mu\nu} &=& \frac{g}{2} (1-\delta_{\alpha\beta})\delta_{\mu\beta}\delta_{\alpha\nu} \\ \nonumber&&+ \frac{\Delta g}{2N} (-1)^{\alpha+\mu} \delta_{\beta,-\alpha}\delta_{\mu,-\nu}.
\end{eqnarray}

With this definition we can write the second order contribution to the total energy as:
\begin{eqnarray}
E^{(2)} &=& \sideset{}{' }\sum_{\{\bk\}} \sum_{\alpha\beta\mu\nu} (G_{\alpha\beta\mu\nu})^2 \frac{n_{\bk_4\nu} n_{\bk_3\mu}(1-n_{\bk_2\beta})(1-n_{\bk_1\alpha}) }{(\bk_4^2 + \bk_3^2-\bk_2^2-\bk_1^2)/2m}.\nonumber\\
\end{eqnarray}

As we are interested in the interaction function, not in the total energy, we evaluate the sums after we vary the energy with respect to the occupation numbers. The result is the following:
\begin{widetext}
\begin{eqnarray}
f^{(2)}_{\alpha\beta} (\bk,\bk') = &-& \left(\frac{g}{2}\right)^2 \Big[ (1-\delta_{\alpha\beta}) \left[I_1(\bk,\bk')+I_2(\bk,\bk')\right]+ \delta_{\alpha\beta} (N-1) I_1(\bk,\bk') \Big] \\ \nonumber
&-& 2\left(\frac{g}{2}\right)\left(\frac{\Delta g}{2N}\right)  \Big[\delta_{\alpha,-\beta} \left[I_1(\bk,\bk')+I_2(\bk,\bk')\right]+ \delta_{\alpha\beta} I_1(\bk,\bk') \Big] \\ \nonumber
&-&  \left(\frac{\Delta g}{2N}\right)^2  \left[ I_1(\bk,\bk')+ \delta_{\alpha,-\beta} \frac{N}{2} I_2(\bk,\bk') \right] ,
\end{eqnarray}
\end{widetext}
where $I_{1,2}(\bk,\bk')$ are the sums:
\begin{eqnarray}
I_1(\bk,\bk') =  \sum_{\bk_1,\bk_2} \frac{n_{\bk_1}-n_{\bk_2}}{(\bk_1^2-\bk_2^2)/2m} \delta_{\bk_1+\bk,\bk_2+\bk'},
\end{eqnarray}

\begin{eqnarray}
I_2(\bk,\bk') =  \sum_{\bk_1,\bk_2} \frac{n_{\bk_1}+n_{\bk_2}}{(2 k_f^2-\bk_1^2-\bk_2^2)/2m} \delta_{\bk_1+\bk_2,\bk+\bk'},
\end{eqnarray}
which can be evaluated as integrals in order to obtain the familiar closed forms:
\begin{eqnarray}
I_1(\bk,\bk') = -\frac{4 k_f m}{(2\pi)^2} \left[ 1 + \frac{1-s^2}{2s} \ln \left(\frac{1+s}{1-s}\right) \right],
\end{eqnarray}

\begin{eqnarray}
I_2(\bk,\bk') = -\frac{8 k_f m}{(2\pi)^2} \left[ 1 - \frac{s}{2} \ln \left(\frac{1+s}{1-s}\right) \right],
\end{eqnarray}
where $s=\sin(\theta/2)$ and $\theta$ is the angle between $\bk$ and $\bk'$, both assumed to be at the Fermi surface.

\begin{figure}[h]
\begin{center}
\includegraphics[width=\linewidth, keepaspectratio]{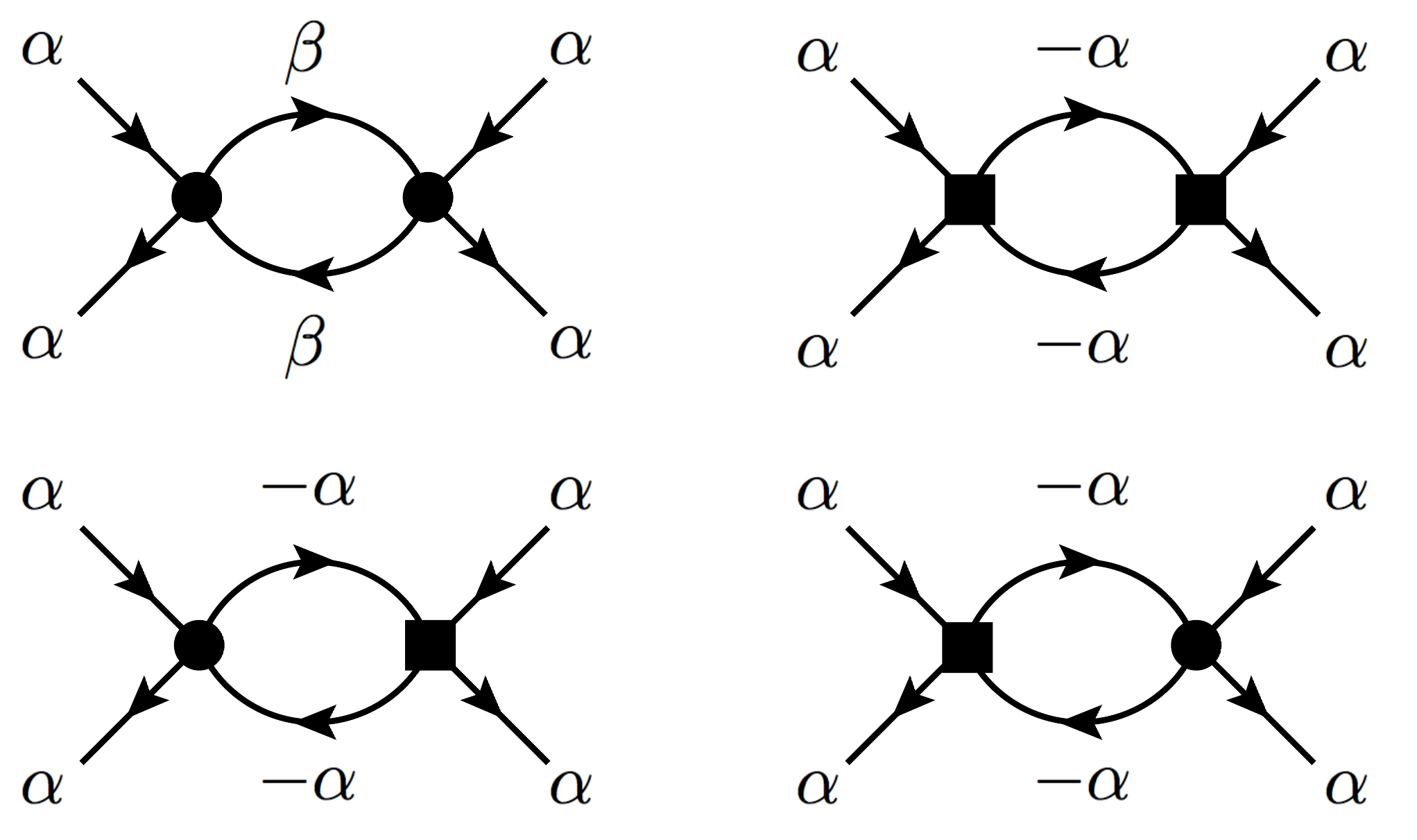}
\caption{Feynmann diagrams related to the interaction function in second order for particles with equal spin.}
\label{DiagAA}
\end{center}
\end{figure}

We can look now at specific cases of the interaction function in second order in the interactions. First, for particles with the same spin:
\begin{eqnarray}
f_{\alpha\alpha} (\bk,\bk') &=& -\left[  \left(\frac{g}{2}\right)^2 (N-1) \right. \\ \nonumber &+& \left. 2 \left(\frac{g}{2}\right)\left(\frac{\Delta g}{2N}\right)+\left(\frac{\Delta g}{2N}\right)^2\right] I_1(\bk,\bk'),
\end{eqnarray}
which we can identify with the diagrams in Fig.~\ref{DiagAA}. Note that the diagram with two circular vertexes, proportional to $\left(\frac{g}{2}\right)^2$, is the only one with internal lines which need to be summed over all the possible spin indexes but $\alpha$, which gives the factor of $N-1$ above. Note also that there is no first order correction in the case of particles with same spin component.

\begin{figure}[h]
\begin{center}
\includegraphics[width=\linewidth, keepaspectratio]{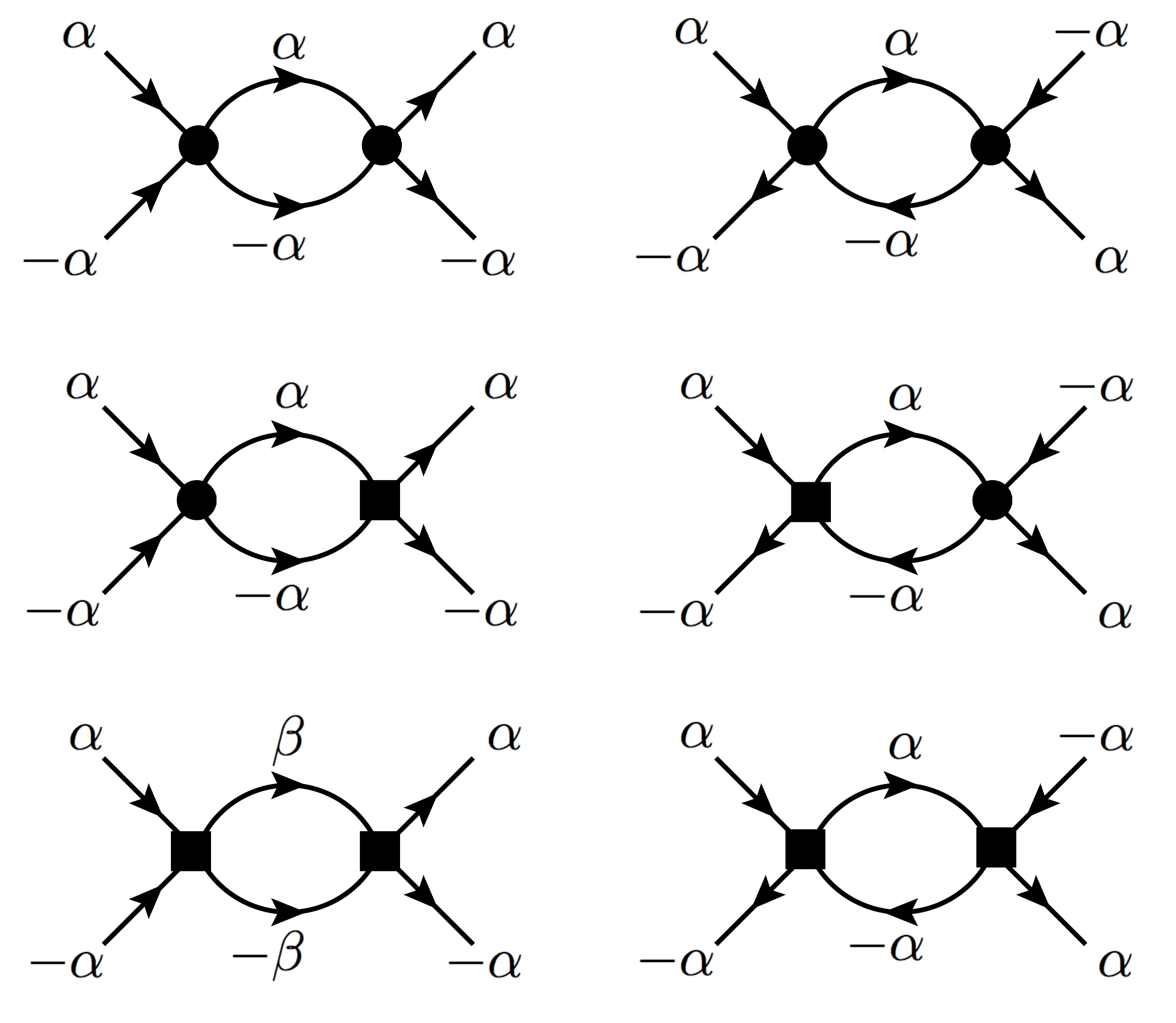}
\caption{Diagrams for the interaction function up to second order for particles with opposite spin components.}
\label{DiagA-A}
\end{center}
\end{figure}

For particles with opposite spin components:
\begin{eqnarray}
f_{\alpha,-\alpha} (\bk,\bk') &=&  \frac{g}{2}+\frac{\Delta g}{2N}\\ \nonumber&-& \left(\frac{g}{2}\right)^2\left[  I_1(\bk,\bk')  + I_2(\bk,\bk')\right] \\\nonumber 
&-&2  \left(\frac{g}{2}\right) \left(\frac{\Delta g}{2N}\right)\left[  I_1(\bk,\bk')  + I_2(\bk,\bk')\right] \\\nonumber
&-&  \left(\frac{\Delta g}{2N}\right)^2\left[I_1(\bk,\bk')+\frac{N}{2} I_2(\bk,\bk') \right],
\end{eqnarray}
which second order terms can be identified with the diagrams in Fig.~\ref{DiagA-A}. Note that this time only one of the diagrams with two square vertexes, proportional to $(\frac{\Delta g}{2N})^2$, have internal lines that need to be summed over, giving rise to the factor of $N/2$ in the last line.

\begin{figure}[h]
\begin{center}
\includegraphics[width=\linewidth, keepaspectratio]{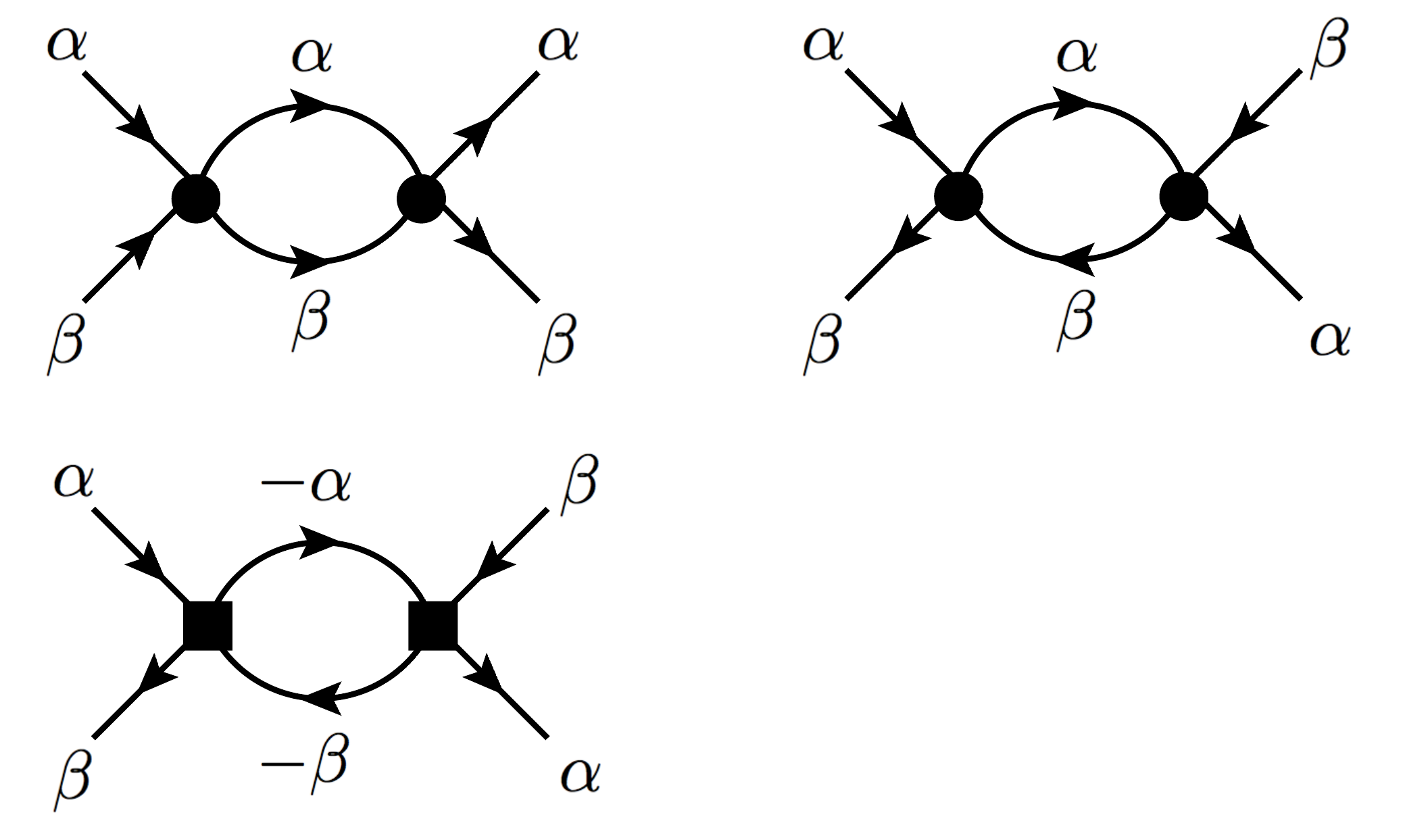}
\caption{Diagrams depicting the interaction function up to second order for particles with spins satisfying $\alpha\neq \pm \beta$.}
\label{DiagAB}
\end{center}
\end{figure}

Finally, for the case of $\alpha\neq \pm\beta$ we find:
\begin{eqnarray}
f_{\alpha,\beta \neq \pm\alpha} (\bk,\bk') &=& \frac{g}{2} - \left(\frac{g}{2}\right)^2\left[  I_1(\bk,\bk')  + I_2(\bk,\bk')\right] 
\\\nonumber  &-& \left(\frac{\Delta g}{2N}\right)^2 I_1(\bk,\bk'),
\end{eqnarray}
related to the diagrams pictured in Fig.~\ref{DiagAB}.

Note that when $\Delta g=g_0-g=0$ we recover the SU(N) results\cite{Yip}. Under this condition $f_{\alpha,-\alpha}=f_{\alpha,\beta\neq \alpha}$ since there are only two FL parameters in the SU(N) case, in contrast to three in the SP(N) case. Due to the different parametrization of the interaction function  in terms of $f_{s,a,\epsilon}(\bk,\bk')$ given the different group properties, in particular the completeness relation, the corrections to the physical observables are going to be different, as will be shown explicitly in the following subsection.

\subsection{Explicit correction to physical quantities}\label{Corr}

Given the results above we can now explicitly write the Fermi liquid parameters $f_{s,a,\epsilon}(\bk,\bk')$. Using the completeness relation for SP(N):
\begin{eqnarray}
\sum_A \Gamma^A_{\alpha\beta}\Gamma^A_{\mu\nu} = \delta_{\alpha\nu} \delta_{\beta\mu}  - \epsilon_{\alpha\mu}\epsilon_{\beta\nu},
\end{eqnarray}
and the fact that the only non-zero interaction functions have the form: $f_{\alpha\beta}(\bk,\bk') \equiv f_{\alpha\beta,\alpha\beta}(\bk,\bk')$, Eq.~\ref{FLPar} can be rewritten as:
\begin{eqnarray}
f_{\alpha,\beta}(\bk,\bk') &=& f_s(\bk,\bk')+ (f_a-f_m)(\bk,\bk')\delta_{\alpha,-\beta} \\\nonumber&+& f_m(\bk,\bk')\delta_{\alpha\beta},
\end{eqnarray}
so we are able to identify:
\begin{eqnarray}
f_s(\bk,\bk')&=& f_{\alpha,\beta\neq\pm\alpha} (\bk,\bk'), \\ \nonumber
f_m(\bk,\bk')&=& (f_{\alpha,\alpha}- f_{\alpha,\beta\neq\pm\alpha})(\bk,\bk'), \\ \nonumber
f_a(\bk,\bk')&=& (f_{\alpha,\alpha}+  f_{\alpha,-\alpha} -  2f_{\alpha,\beta\neq\pm\alpha})(\bk,\bk').
\end{eqnarray} 

\begin{figure}[t]
\begin{center}
\includegraphics[width=\linewidth, keepaspectratio]{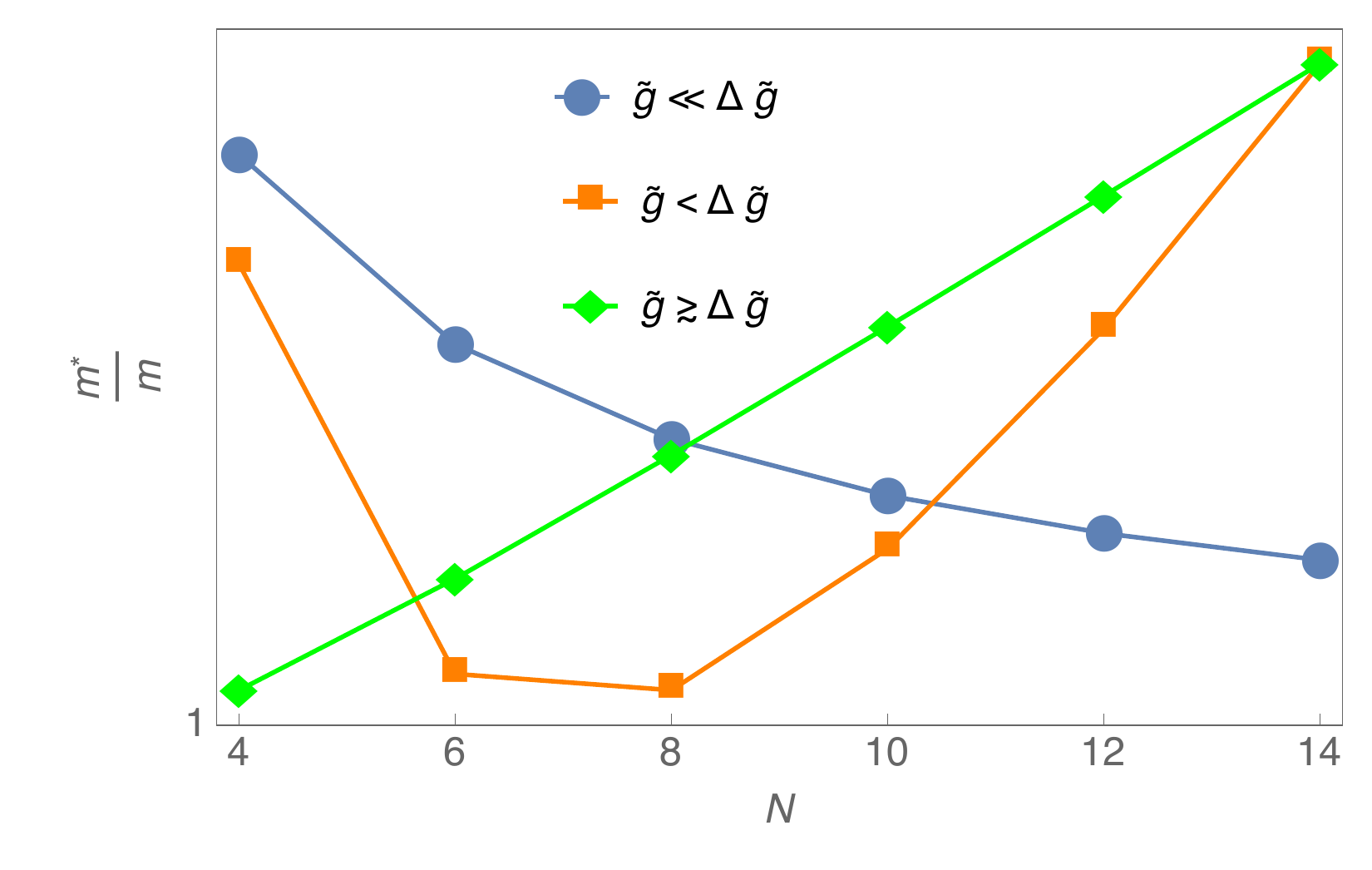}
\caption{Renormalization of the effective mass as a function of N. Note the different qualitative behaviour for different parameter regimes. The curves above have different ranges and parameters. The blue circles were plotted for $\tilde{g}=0.005$ and $\Delta \tilde{g} = 0.5$ with the curve ranging up to 1.08; the orange squares used $\tilde{g}=0.1$ and $\Delta \tilde{g} = 0.5$  and range up to 1.125; the green diamonds were plotted for $\tilde{g}=0.1$ and $\Delta \tilde{g} =0.1$, with the curve ranging up to 1.10. The lines are guides to the eye.} 
\label{Mass}
\end{center}
\end{figure}

The effect of interactions on physical quantities appear in terms of $f_{s,m,a}(\bk,\bk')$, averaged over the Fermi surface. The calculation involves then averages of combinations of $I_1(\bk,\bk')$ and $I_2(\bk,\bk')$ over the Fermi surface, what leads to well known integrals. The effective mass reads:
\begin{eqnarray}
\frac{m^*}{m} = 1+ \frac{16}{15}\Bigg[ && \!\!\!\!\!\left(\frac{\tilde{g}}{2}\right)^2  \left((5+N) \ln2-5+2N\right) \nonumber\\
&+& 2 \left(\frac{\tilde{g}}{2}\right) \left(\frac{\Delta \tilde{g}}{2N}\right) (7\ln2-1)\nonumber\\
&+& \left(\frac{\Delta \tilde{g}}{2N}\right)^2 N (7\ln2-1)\Bigg]
\end{eqnarray}
and the compressibility has a similar form:
\begin{eqnarray}
\left(\frac{u^{*}}{u}\right)^2 = 1&+& \frac{N}{2}\left[(N-1)\left(\frac{\tilde{g}}{2}\right)^2 \frac{8}{3}(2\ln2+1) \right.\nonumber\\ &+&
2 \left(\frac{\tilde{g}}{2}\right) \left(\frac{\Delta \tilde{g}}{2N}\right) \frac{16}{3} (\ln2 +2)\\ \nonumber&+&\left. \left(\frac{\Delta \tilde{g}}{2N}\right)^2 \frac{16}{3} \big((2-N)\ln2 + N +1\big)\right].
\end{eqnarray}
Here we defined the dimensionless quantities $\tilde{g} = \rho(E_f) g$ and $\Delta\tilde{g} = \rho(E_f) \Delta g$. By inspection one can see that for $N>2$ the corrections to the effective mass and compressibility are always positive, even in the case $\Delta \tilde{g}<0$, leading to an enhancement of both quantities due to interactions. Interestingly enough, the behaviour of the enhancement of the effective mass and compressibility as a function of N varies for different parameter regions, as shown in Fig.~\ref{Mass}. We focus on the small $\tilde{g}$ and $\Delta \tilde{g}$ parameter region since the calculation is perturbative in these parameters. For $ \tilde{g}\ll \Delta \tilde{g}$ the enhancement decreases as a function of N, while for $\tilde{g} \gtrsim \Delta \tilde{g}$  the enhancement increases as a function of N. For intermediate regimes a non-monotonic dependence on N can be observed. These qualitative features are valid for both $\Delta \tilde{g}>0$ and $\Delta \tilde{g} <0$.

Concerning the correction to the magnetic susceptibility, we can look at the Wilson ratio, as a measure of the enhancement of the susceptibility due to exchange interactions:
\begin{eqnarray}
W= \frac{\chi_{G,P}^*/m^*}{\chi_{G,P}/m}=\frac{1}{1+   \overline{F_a(\theta)}}.
\end{eqnarray}
which takes the explicit form:
\begin{eqnarray}
W= 1 &-& \frac{\Delta \tilde{g}}{2N} -  8 \left(\frac{\tilde{g}}{2}\right) \left(\frac{\Delta \tilde{g}}{2N}\right) -\left(\frac{\Delta \tilde{g}}{2N}\right)^2 \frac{8N}{3}(1-\ln 2).\nonumber\\
\end{eqnarray}
Considering the Wilson ratio in first order in the interactions, there can be an instability for $\Delta \tilde{g}>0$ at $\Delta \tilde{g}/2N=1$. Note that there is an N-dependence for the instability in first oder, different from the SU(N) case\cite{Yip}. The larger the N, the harder is for the system to reach the instability, or the larger the interaction needed for the magnetic order to set in. 

The second order contribution to the Wilson ratio puts the system closer to an instability, assuming that $\tilde{g}>0$ as we have originally repulsive interactions. This is in contrast with the findings for SU(N), since for $N>2$ the second order corrections always take the system away from a magnetic instability\cite{Yip}.


\begin{figure}[t]
\begin{center}
\includegraphics[width=\linewidth, keepaspectratio]{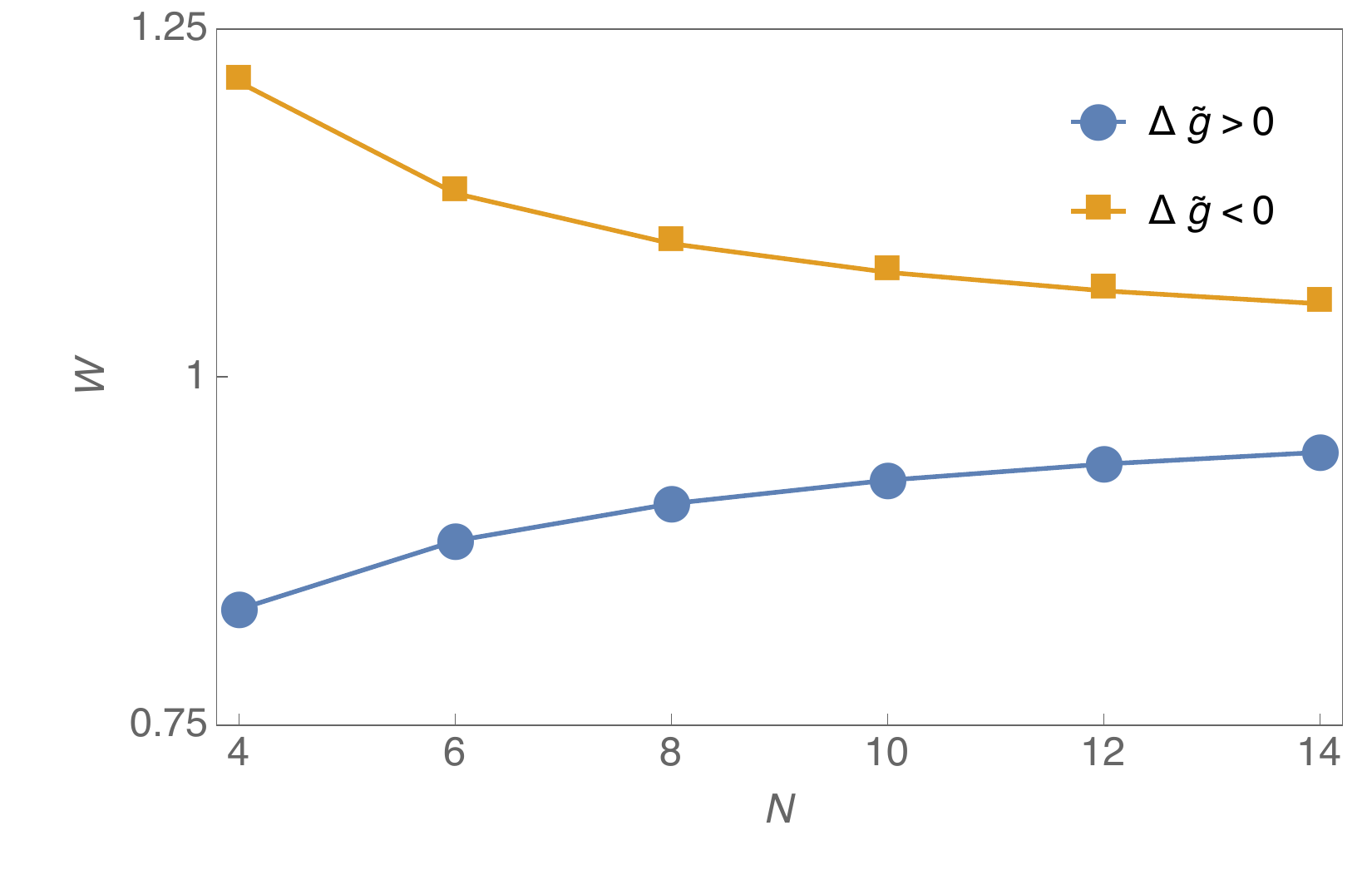}
\caption{Wilson ratio as a function of N. For this plot the parameters used are $\tilde{g}=0.5$ and $\Delta \tilde{g}=\pm 0.5$. } 
\label{Susceptibility}
\end{center}
\end{figure}

Here we would like to notice that one cannot benchmark these results with the SU(N) case by directly taking the limit $\Delta \tilde{g}=0$ since we are using a different parametrization (given the different completeness relations for the generators of the different symmetry groups). The benchmarking needs to go one step back, eliminating the parameters $f_a(\theta)$ from the evaluation of the corrections above and taking $\Delta \tilde{g} =0$, in which case the SU(N) results are recovered\cite{Yip}. We also note that the limit with $N=2$ cannot be directly taken since in this case $\tilde{g}_0$ is the only one scattering channel. Again one needs to go one step back and make $\tilde{g}=0$ in order to recover the SU(2) results.

\section{Discussion and Conclusion}\label{Con}

\subsection{Experimental realization and verification}\label{ExpRea}

We now discuss what are the candidates for the realization of symplectic symmetry within ultra-cold fermionic systems. We start restricting ourselves to atoms whose scattering properties can be well described by contact interaction in the limit of ultra-cold temperatures, so we neglect atoms with sizeable dipole-dipole interactions\cite{MaiT}. We also focus on stable or long lived isotopes which can be actually handled in an experimental setup. These restrictions eliminate some of the transition metals, lanthanides and actinides which have large dipole-dipole interactions and elements heavier than Pb. We now go over the different families in the periodic table, exploring what are the possibilities to realize enlarged symmetries. The discussion below is summarized in Table \ref{TabEl}.

\emph{Alkali Metals}: These have the outer electronic shell with configuration $ns^1$ (here $n$ is the principal quantum number). The total electronic angular momentum is $J=1/2$. As we are interested in atoms which are fermions, we need to look for isotopes with integer nuclear spin. This leaves us with $^2$H, $^6$Li and $^{40}$K, with nuclear spin $I$ equal to 1, 1 and 4, respectively. Due to the hyperfine interaction these combine into a total hyperfine spin $f$ equal to 1/2, 1/2 and 9/2. The first two cases are trivial in the sense that SP(2) is isomorphic to SU(2). In conclusion, $^{40}$K is the promising candidate among the alkali metals.

\emph{Alkaline-Earth Metals}: These have a full outer electronic shell with configuration $ns^2$. In this case $J=0$ and the electronic degrees of freedom are therefore decoupled from the (possibly non-zero) nuclear spin. In order to obtain a fermionic isotope we need a nucleus with half-integer spin. Now we have the following options: $^9$Be, $^{25}$Mg, $^{41,43}$Ca, $^{87}$Sr and $^{135,137}$Ba, with nuclear spins ranging from 3/2 to 9/2, as summarized in Table \ref{TabEl}. These are known to realize SU(N) symmetry, so in principle one could realize enlarged symmetries ranging from SU(4) up to SU(10) with alkaline-earth atoms.

\emph{Transition Metals}: One would expect that there would be also some candidates amongst the transition metals. For the family IB, with $ns^1$ electronic configuration, one would need a nucleus with an even number of nucleons, but nature does not provide us with stable isotopes of this kind. The atoms in family IIB have a full electronic shell of s-character, so we look for isotopes with half-integer nuclear spin. There are several: $^{67}$Zn, $^{111,113}$Cd and $^{199,201}$Hg, with nuclear spins ranging from 1/2 to 5/2. Since these have a full electronic shell, they also realize SU(N) symmetry. Other transition metals have large dipole-dipole interactions or do not have stable fermionic isotopes.

\emph{Lanthanides}: Amongst the Lanthanides Yb is one of the few elements with no dipole-dipole interaction due to its complete electronic shell. There are two isotopes which are fermions: $^{171,173}$Yb, with nuclear spin equal to 1/2 and 5/2, respectively. These would realize SU(2) and SU(6) symmetries, respectively.

\emph{Families IIIA-VIIA}: Elements in the families VA-VIIA have a substantial multipolar character, so we are not going to consider them. Elements in the families IIIA have an odd number of electrons, therefore we should look for integer nuclear spin isotopes. The only stable isotope is $^{10}$B with nuclear spin equal to 3. Elements of the families IVA have an even number of electrons, therefore we are interested in isotopes with half-integer spin. There are several isotopes available in nature: $^{13}$C, $^{29}$Si, $^{73}$Ge, $^{115,117}$Sn and $^{207}$Pb with nuclear spins equal to 1/2, with the exception of $^{73}$Ge which has nuclear spin equal to 9/2.

\emph{Noble Gases}: These have a complete electronic shell, so $J=0$ and we should look for isotopes with half-integer nuclear spin. The stable isotopes are $^{3}$He, $^{21}$Ne, $^{83}$Kr, $^{129,131}$Xe, with hyperfine spins ranging between 1/2 and 9/2. These also realize SU(N) symmetry.


From the analysis above we can conclude that nature is quite unfair towards the realization of symplectic symmetry. The most interesting case would be to have an isotope with hyperfine spin $f=3/2$ and electronic angular momentum $J\neq 0$, which would not require fine-tuning of the scattering channels. In this case there are only two interaction channels satisfying $g_0\neq g_2$. Unfortunately there is no such isotope (at least not on its ground state and without substantial dipole-dipole interaction). The only $f=3/2$ cases are within the elements with $J=0$, so they actually realize the larger SU(4) symmetry and not SP(4) symmetry.  Note that the discussion in terms of the interaction strengths in different channels, $g_F$, is analogous to the discussion in terms of scattering lengths, $a_F$, since these are related by the identity $\rho(E_f) g_F = 2 k_f a_F/\pi \hbar$, where as before $k_f$ is the Fermi momentum and $\rho(E_f)$ the density of states at the Fermi level.

\renewcommand{\arraystretch}{1.5}

\begin{table*}[t]
  \begin{tabular}{ | c | c | c | c | c | c |}
    \hline
\begin{tabular}{c} Electronic\\Configuration \end{tabular} & Family & Isotope & Nuclear Spin & Hyperfine Spin & Symmetry\\ \hline
\multirow{12}{*}{J=0} &
\multirow{5}{*}
	{\begin{tabular}{c}Alkaline-Earth \\Metals\end{tabular}}
	  & $^9$Be&3/2&3/2&SU(4)\\ \cline{3-6}
	  && $^{25}$Mg&5/2&5/2&SU(6)\\ \cline{3-6}
	  && $^{41,43}$Ca&7/2&7/2&SU(8)\\ \cline{3-6}
	  && $^{87}$Sr&9/2&9/2&SU(10)\\ \cline{3-6}
	  && $^{135,137}$Ba&3/2&3/2&SU(4)\\ \cline{2-6}
&\multirow{1}{*}{Lanthanides}
	  & $^{173}$Yb&5/2&5/2&SU(6)\\ \cline{2-6}
&\multirow{2}{*}{Family IIB}
	  & $^{67}$Zn&5/2&5/2&SU(6)\\ \cline{3-6}
	  && $^{201}$Hg&5/2&3/2&SU(4)\\ \cline{2-6}
 &\multirow{3}{*}{Noble Gases}
	  & $^{21}$Ne&3/2&3/2&SU(4)\\ \cline{3-6}
	  && $^{83}$Kr&9/2&9/2&SU(10) \\ \cline{3-6}
  	  && $^{131}$Xe&3/2&3/2&SU(4)\\  
    \hline
     \multirow{2}{*}{J=1/2} &
 \multirow{1}{*}{Alkali Metals}
	 & $^{40}$K&4&9/2&SP(10)$^*$\\  \cline{2-6}
	 &Family IIIA&$^{10}$B &3&5/2&SP(6)$^*$\\  \hline
	      \multirow{1}{*}{J=1} &
 \multirow{1}{*}{Family IVA}
	 & $^{73}$Ge&9/2&7/2&SP(8)$^*$\\  \cline{3-6} \hline
  \end{tabular}
\caption{List of stable or long-lived isotopes with hyperfine spin larger than 1/2, and the respective symmetry they realize. * Requires fine-tuning.}
\label{TabEl}
\end{table*}

There are elements with larger hyperfine spins which have $J\neq 0$. These are: $^{40}$K, $^{10}$B and $^{73}$Ge, and can realize SP(N) symmetry in case the scattering lengths are fine tuned such that $a_0\neq a_{F>0}$.  We focus on $^{40}$K, the only one amogst these that was already taken to ultra-low temperatures. Interestingly, it has a very large hyperfine spin ($f=9/2$) so the associated symmetry is SP(10). Note that the symmetry will be present only if the scattering lengths $a_{F=2,4,6,8}$ are all made equal. To fine tune four parameters in the system looks like a challenge, but in fact $^{40}$K is already very close to naturally satisfy this condition. From Krauser et al.\cite{Kra}, one can see that the scattering lengths for $F>0$ are the same within about $12\%$ 
($a_0\sim120$, 
$a_2=147.83$, 
$a_4= 161.11$,
$a_6=166.00$,
$a_8= 168.53$ in units of the Bohr radius). Given the fact that we are naturally close to the fine-tuned point with all $g_{F>0}$ equal, it might be interesting to explore how one could tune the system towards better satisfying this condition, perhaps by the use of Feshbach resonances.

As already discussed in Sec.~\ref{SPN}, for SP(N) systems we cannot simply initially load it with a few of the states in order to realize a smaller SP(n$<$N) symmetry, as can be done in the SU(N) case. If we load the system with one flavour and its complement, it can scatter to another flavour and its component, $\alpha,-\alpha\rightarrow \beta,-\beta$, as sketched in Fig.~\ref{SUNSPN} e). If that was possible, one could take $^{40}$K and load the system only with the states $\{\pm1/2,\pm3/2\}$, realizing SP(4) symmetry. In this direction, one could think on engineering a way to block the scattering to other states.

Another possibility to realize SP(N) symmetry would be to think on the other way around: it might be possible to detune the $F=0$ channel away from the remaining channels in an isotope which has SU(N) symmetry so we are able to break it down to SP(N) symmetry. In principle this can be achieved by connecting the low lying states with excited states by external fields.

One could also explore the atoms we neglected above, with strong dipole-dipole interactions, by tuning them with Feshbach resonances such that the contact interactions are much stronger than the dipole-dipole interactions. One of the atoms with the strongest dipole-dipole interaction is Dy and there are two fermionic isotopes: $^{161,163}$Dy, both with nuclear spin equal to 5/2. Another atom with significant dipolar character is Er, which has only one stable fermionic isotope, $^{167}$Er. Recently Er and Dy have been shown to display a very dense Feshbach spectrum with signatures of chaotic behaviour \cite{Mai}. This suggests that one could scan the system as a function of magnetic field in order to find a value which gives the suitable scattering lengths following $a_0\neq a_{F\neq 0}$, in similar fashion to what was done by Lahaye at al.\cite{Lah}. $^{53}$Cr is another dipolar isotope which was already brought to a degenerate state \cite{Nay}.

In order to verify the presence of SP(N) symmetry one could do more than simply measuring the scattering lengths in different channels. One can track the evolution of the occupation number of each spin component. For the SU(N) case each component is preserved independently, so there should be no changes in $n_\alpha$ over time. For SP(N) the color-magnetizations $m_\alpha=n_\alpha - n_{-\alpha}$ are the conserved quantities, so even though $n_{\alpha}$ can change over time, this specific difference does not. This can be verified by Stern-Gerlach experiments, similar to the one performed by Krauser et al.\cite{Kra}.

\subsection{Realization of Physics in Higher Dimensions}\label{highd}

Symmetries dictate the kinds of quantum states, or particle types, that can exist. If we are in isotropic space in d-dimensions, the physics should be invariant under SO(d) transformations, for massive particles with positive definite energy\cite{Wig,Wei}. The different types of particles that exist correspond to the different irreducible projective representations of SO(d), or to the distinct irreducible representations of its covering group, the Spin(d) group\cite{Moo,Law}. The representations of Spin(d) which are not representations of SO(d) are spinor representations, associated with fermionic particles. Interestingly enough, for low dimensions there is a series of interesting isomorphisms, in particular, Spin(3) $\cong$ SU(2) $\cong$ SP(2) and Spin(5) $\cong$ SP(4)\cite{Moo,Law}. These isomorphisms suggest the following correspondence: if we start with a 3-dimensional world, in which case the physics is invariant under SO(3) rotations, we can look for the irreducible representations  of SO(3) and we find that they correspond to all integer angular momentum states. If we now look at representations of SU(2), its double-cover, we find extra representations which are related to half-integer spin particles, or fermions. Interestingly enough SP(2) has 3 generators, corresponding to the three different spin components, which couple, correspondingly, to the three different magnetic field components in three dimensional space. In the same fashion, if we are in $d=5$, the double-cover of SO(5) is Spin(5) $\cong$ SP(4). If we are able to realize a system with effective SP(4) symmetry, in particular in the fundamental representation (the smallest faithful representation), we are in fact realizing the analogous of spin-1/2 particles in 3-dimensions, but now in 5- dimensions. SP(4) has $10$ generators, corresponding to 10 different ``spin components". Note that in five dimensions the number of magnetic field components is also 10. Defining a generalized magnetic field as the possible antisymmetric pair-wise combinations of the electromagnetic tensor components $F_{ij}$, with $i,j=1,...,d$, we have in total $d(d-1)/2$ magnetic field components.

One could think of these particles confined to three dimensions in the same way as we talk about SU(2) spin-1/2 particles in one- or two-dimensional systems. Even though it is not clear how to manipulate the fictitious magnetic field in higher dimensions one could still measure fluctuations of the system and determine its response functions by the use of the fluctuation-dissipation theorem. This is an interesting direction for future work.




\subsection{Final Remarks}

In conclusion we have reviewed the problem of interacting fermions with SP(N) symmetry within cold atoms and contrasted its behaviour with the SU(N) scenario. We characterized the main properties of the Fermi Liquid state: effective mass, compressibility and magnetic susceptibility. We find that both the effective mass and inverse compressibility are enhanced in the presence of interactions following SP(N) symmetry. The magnetic susceptibility can be either enhanced or suppressed, depending on the sign of the detuning parameter $\Delta \tilde{g}=\rho(E_f) (g_0-g)$. We conclude discussing what are the possible routes to realize SP(N) symmetry within cold atoms, which apparently always requires some degree of fine-tunning, setting up an interesting challenge for experimentalists. The correspondence of SP(4) systems to physics in 5-dimensions is a fascinating direction for future work.





\begin{acknowledgements}
I thank Ana Maria Rey, Tilman Esslinger, Gordon Baym, Carlos Sa de Melo and Gregory W. Moore for illuminating discussions. I also thank Sungkip Yip for bringing to my attention his work on the SU(N) version and Rémi Desbuquois for very insightful discussions on the possible experimental paths to realize SP(N) symmetry. I also thank Manfred Sigrist and Yehua Liu for carefully reading and commenting on a preliminary version of this manuscript. This work was performed in part at the Aspen Center for Physics, which is supported by National Science Foundation grant PHY-1066293. This work was also supported by Dr. Max R\"{o}ssler, the Walter Haefner Foundation and the ETH Zurich Foundation. 
\end{acknowledgements}

\appendix
\section{Generators of the symplectic group}\label{AppGen}

The definition of the generators for the symplectic-N generalization in reference \cite{Fli08} is different from the one we use in this manuscript. Here we define:
\begin{eqnarray}\label{DefGen}
S_{\alpha\beta} = \Psi_\alpha^\dagger \Psi_\beta+ (-1)^{\alpha+\beta} \Psi_{-\beta}^\dagger \Psi_{-\alpha},
\end{eqnarray}
with $\alpha$ and $\beta$ ranging from $-f$ to $f$ for $N=2f+1$.

We can show that the above operator form in fact gives a set of generators of SP(N) by looking at the properties of their matrix form in a specific basis.  There are $N(N+1)/2$ linearly independent $S_{\alpha\beta}$ which are related to $N \times N$ matrices $M_i$, $i=1,...,N(N+1)/2$, which follow the symplectic condition:
\begin{eqnarray}
M_i^T \Omega + \Omega M_i =0,
\end{eqnarray} 
where $\Omega$ is an antisymmetric matrix.

It can be shown that these are the generators of the symplectic group by using the explicit  matrix forms, in the basis $(\Psi_{3/2}, \Psi_{1/2},\Psi_{-1/2},\Psi_{-3/2})$: 
\begin{eqnarray}
[S_{\alpha\beta}]_{mn} &=& \delta_{m, s-\alpha+1}\delta_{n,s-\beta+1} \\ \nonumber&+& (-1)^{\alpha+\beta} \delta_{m,s+\beta+1}\delta_{n,s+\alpha+1},
\end{eqnarray} 
and the antisymmetric form $\Omega=AntiDiag[1,-1,1,-1,...]$:
\begin{eqnarray}
[\Omega]_{mn} = (-1)^m \delta_{m, 2s-n+2}.
\end{eqnarray} 
Verifying the symplectic condition:
\begin{eqnarray}
&&[S_{\alpha\beta}^T]_{mn} [\Omega]_{np} + [\Omega]_{mn} [S_{\alpha\beta}]_{np} = 0,
\end{eqnarray} 
using the fact that $[S_{\alpha\beta}^T]_{mn}=[S_{\alpha\beta}]_{nm}$ and the explicit matrix forms given above we find:
\begin{eqnarray}
&&  (-1)^{n+1}  \delta_{n,s-\alpha+1} \delta_{m, s-\beta+1}   \delta_{n,2s-p+2}\\ \nonumber
&+& 
 (-1)^{\alpha+\beta+n+1} \delta_{n, s+\beta+1}\delta_{m,s+\alpha+1} \delta_{n,2s-p+2} \\ \nonumber
&+&  (-1)^{m+1} \delta_{m, 2s-n+2} \delta_{n, s-\alpha+1}\delta_{p, s-\beta+1}\\ \nonumber
&+& (-1)^{\alpha+\beta + m +1} \delta_{m, 2s-n+2} \delta_{n,s+\beta+1}\delta_{p,s+\alpha+1}=0,
\end{eqnarray} 
which is zero if we are working with fermions and $s$, $\alpha$ and $\beta$ are half-integers. 

Here we note that we would get the same matrix form for the generators as in Flint et al. \cite{Fli08} if we use a different basis: $(\Psi_{3/2}, \Psi_{-1/2},\Psi_{1/2},\Psi_{-3/2})$, in which case the antisymmetric matrix has a different form: $\tilde{\Omega}=AntiDiag[1,1,1,...,-1,-1,-1,...]$. In Flint et al. \cite{Fli08} the basis used is $(\Psi_{3/2}, \Psi_{1/2},\Psi_{-1/2},\Psi_{-3/2})$ with the same measure.

Note that the generators as presented above are traceless but not properly orthonormalized. For the development of the FL theory it is convenient to work with generators which are orthonormal. This can be achieved by rescaling these generators as $\tilde{S}_{\alpha\beta}=S_{\alpha\beta}/\sqrt{2}$ and combining them as follows:

A) Generators of the form $\tilde{S}_{\alpha\alpha}$, with both indexes equal. Given the relation $\tilde{S}_{\alpha\alpha}=-\tilde{S}_{-\alpha-\alpha}$, we need to consider only the generators with positive indexes. There are $N/2$ of those and these are already properly orthonormalized such that $Tr[\tilde{S}_{\alpha\alpha} \tilde{S}_{\beta\beta}]=\delta_{\alpha\beta}$.

B)  There are also $N$ linearly independent generators with opposite indexes as $\tilde{S}_{\alpha-\alpha}$. To guarantee orthonormality these should be reorganized as $(\tilde{S}_{\alpha-\alpha} + \tilde{S}_{-\alpha\alpha})/2$ and $(\tilde{S}_{\alpha-\alpha} - \tilde{S}_{-\alpha\alpha})/2i$.

C) The missing generators have the form $\tilde{S}_{\alpha\beta}$ with $\alpha\neq \pm \beta$. Given again the relation $\tilde{S}_{\alpha\beta}=(-1)^{\alpha+\beta}\tilde{S}_{-\beta-\alpha}$, if both indexes are positive there are $\frac{N}{2}\left(\frac{N}{2}-1\right)$, and we do not need to consider the generators with both negative indexes. If one index is positive and the other negative, it is linearly dependent of another generator of the same form, so again there are $\frac{N}{2}\left(\frac{N}{2}-1\right)$. This totals $N\left(\frac{N}{2}-1\right)$ generators of the form $\tilde{S}_{\alpha\beta}$ with $\alpha\neq \pm \beta$. These should be combined in such a way that they are all summed but one, which is subtracted, what leads to $N\left(\frac{N}{2}-1\right)$ independent combinations.

Note that the total number of generators is still $N(N+1)/2$. In the main text we refer to this set of properly orthonormalized generators as $\Gamma^A$, such that $Tr[\Gamma^A\Gamma^B]=\delta_{AB}$, without writing them explicitly.

\section{Parametrization of the interaction function for the SP(N) FL}\label{AppSym}

In order to understand the parametrization of the interaction function, we can look at the total change in energy due to a change in the occupation number $\delta n_{\alpha\beta}(\bk)$:
\begin{eqnarray}
\delta E &=&\sum_{\bk, \alpha\beta} \delta\epsilon_{\alpha\beta}(\bk) \delta n_{\beta\alpha}(\bk)\\ \nonumber
&=& \sum_{\bk, \bk',\alpha\beta\mu\nu} f_{\alpha\mu,\beta\nu}(\bk,\bk' )\delta n_{\nu\mu}(\bk') \delta n_{\beta\alpha}(\bk).
\end{eqnarray}
We can now go back to the operator form in order to analyze the symmetries more explicitly, identifying $n_{\alpha\beta}(\bk) = \langle\Psi_{\bk\beta}^\dagger\Psi_{\bk\alpha}\rangle$, we have that the change in the total energy has the form
\begin{eqnarray}
\sim  \Psi_{\bk\alpha}^\dagger \Psi_{\bk' \mu}^\dagger f_{\alpha\mu,\beta\nu}(\bk,\bk' ) \Psi_{\bk\beta} \Psi_{\bk' \nu},
\end{eqnarray}
before taking the averages, with the sum over repeated indexes implied in the equation above and in the following. If there is a unitary symmetry group whose transformations are denoted by $U$, we can rotate the operators in spin space such that this form is invariant. We can write $\Psi_{\alpha} = \sum_a U_{\alpha a} \Psi_a$ and rewrite the equation above as:
\begin{eqnarray}
\sim \Psi_{\bk a}^\dagger U_{a\alpha}^\dagger \Psi_{\bk' c}^\dagger U_{c\gamma}^\dagger f_{\alpha\gamma,\beta\delta}(\bk,\bk' )  U_{\beta b}\Psi_{\bk b}   U_{\delta d}\Psi_{\bk' d},
\end{eqnarray}
so in order for this term to be invariant:
\begin{eqnarray}
U_{a\alpha}^\dagger U_{c\gamma}^\dagger f_{\alpha\gamma,\beta\delta}(\bk,\bk' )  U_{\beta b}  U_{\delta d} = f_{ac,bd}(\bk,\bk' ).
\end{eqnarray}
If the transformation is unitary $U U^\dagger= I$, the identity above can be satisfied if $f_{\alpha\gamma,\beta\delta} = \delta_{\alpha\beta}\delta_{\gamma\delta}$:
\begin{eqnarray}
U_{a\alpha}^\dagger U_{c\gamma}^\dagger( \delta_{\alpha\beta}\delta_{\gamma\delta}) U_{\beta b}  U_{\delta d} &=& U_{a\alpha}^\dagger U_{c\gamma}^\dagger  U_{\alpha b}  U_{\gamma d}\\ \nonumber
&=& (U^\dagger U )_{ab} (U^\dagger U )_{cd} \\ \nonumber &=& \delta_{ab}\delta_{cd},
\end{eqnarray}
and also for $f_{\alpha\gamma,\beta\delta} = \delta_{\alpha\delta}\delta_{\gamma\beta}$:
\begin{eqnarray}
U_{a\alpha}^\dagger U_{c\gamma}^\dagger( \delta_{\alpha\delta}\delta_{\gamma\beta}) U_{\beta b}  U_{\delta d} &=& U_{a\alpha}^\dagger U_{c\gamma}^\dagger  U_{\gamma b}  U_{\alpha d}\\ \nonumber
&=& (U^\dagger U )_{ad} (U^\dagger U )_{cb} \\ \nonumber &=& \delta_{ad}\delta_{cb}.
\end{eqnarray}
For the SU(N) case these are the only possible constructions based on the unitarity of the transformations.

In the SP(N) case, given the symplectic condition for the transformations $\epsilon U^{T} = U^\dagger\epsilon$, there is one more possibility $f_{\alpha\gamma,\beta\delta} = \epsilon_{\alpha\gamma}\epsilon_{\beta\delta}$:
\begin{eqnarray}
U_{a\alpha}^\dagger U_{c\gamma}^\dagger(\epsilon_{\alpha\gamma}\epsilon_{\beta\delta}) U_{\beta b}  U_{\delta d} &=& U_{a\alpha}^\dagger\epsilon_{\alpha\gamma} U_{\gamma c}^* U_{b\beta }^T  \epsilon_{\beta\delta} U_{\delta d} \\ \nonumber
&=& \epsilon_{a\alpha}U_{\alpha\gamma}^T U_{\gamma c}^*   \epsilon_{b\beta}U_{ \beta\delta}^\dagger U_{\delta d}\\ \nonumber
&=& \epsilon_{a\alpha}\delta_{\alpha c}   \epsilon_{b\beta}\delta_{\beta d}\\ \nonumber
&=& \epsilon_{ac}\epsilon_{bd}.
\end{eqnarray}

\section{Some properties of the Clebsch-Gordan coefficients}\label{AppCGC}
\begin{widetext}
Below are some useful properties of the CGC $ \langle j_1 m_1, j_2 m_2 | J m \rangle  =  \langle J m | j_1 m_1, j_2 m_2 \rangle $ used in the main text:
\begin{eqnarray}\label{CGC1}
 \langle j_1 m_1, j_2 m_2 | J m \rangle  =  (-1 )^{j_1+j_2-J }  \langle j_1 -m_1, j_2 -m_2 | J -m \rangle,
\end{eqnarray}
\begin{eqnarray}\label{CGC2}
 \langle j_1 m_1, j_2 m_2 | J m \rangle  =  (-1 )^{j_1+j_2-J }  \langle j_2 m_2, j_1 m_1 | J m \rangle,
\end{eqnarray}
\begin{eqnarray}\label{CGC0}
 \langle j_1 m_1, j_2 m_2 | 0 0  \rangle  =  \delta_{j_1,j_2} \delta_{m_1-m_2}\frac{(-1 )^{j_1-m_1 }}{\sqrt{2 j_1+1}},
\end{eqnarray}
and the orthogonality relations:
\begin{eqnarray}\label{CGCOrt1}
\sum_{F,M}  \langle f \alpha, f \beta | F M \rangle \langle F M | f \mu, f \nu\rangle =  \langle f \alpha, f \beta  | f \mu, f \nu\rangle =\delta_{\alpha \mu}\delta_{\beta \nu},
\end{eqnarray}
\begin{eqnarray}\label{CGCOrt2}
\sum_{\alpha,\beta} \langle F M  |  f \alpha, f \beta \rangle \langle  f \alpha, f \beta | F'M'\rangle =   \langle F M  | F'M'\rangle  =\delta_{F, F'}\delta_{M, M}.
\end{eqnarray}
\end{widetext}


\end{document}